\documentclass[onecolumn,showpacs,preprintnumbers,amsmath,amssymb,]{revtex4}
\usepackage[latin9]{inputenc}
\usepackage{amsmath}
\usepackage{amssymb}
\usepackage{graphicx}
\usepackage{color}
\usepackage{slashed}
\usepackage{mathrsfs}
\usepackage{amsbsy}
\usepackage{bm}
\usepackage{float}
\usepackage{epstopdf}
\usepackage[unicode=true,
 bookmarks=true,bookmarksnumbered=false,bookmarksopen=false,
 breaklinks=false,pdfborder={0 0 1},backref=false,colorlinks=false]
 {hyperref}

\makeatletter

\@ifundefined{textcolor}{}
{%
 \definecolor{BLACK}{gray}{0}
 \definecolor{WHITE}{gray}{1}
 \definecolor{RED}{rgb}{1,0,0}
 \definecolor{GREEN}{rgb}{0,1,0}
 \definecolor{BLUE}{rgb}{0,0,1}
 \definecolor{CYAN}{cmyk}{1,0,0,0}
 \definecolor{MAGENTA}{cmyk}{0,1,0,0}
 \definecolor{YELLOW}{cmyk}{0,0,1,0}

}

\makeatother

\begin{document}

\title{A Two-Gauge Field Model for Magnetoelectric Boundaries}

\author{F.A. Barone$^{1}$}
\email{fbarone@unifei.edu.br}

\author{H.L. Oliveira$^{2}$}
\email{helderluiz10@unifei.edu.br}

\author{J. P. Ferreira}
\email{ferreira\_jp@unifei.edu.br}

\affiliation{IFQ - Universidade Federal de Itajub\'a, Av. BPS 1303, Pinheirinho, Caixa Postal 50, 37500-903, Itajub\'a, MG, Brazil}


\begin{abstract}
This work introduces a field-theoretical model designed to simulate the presence of material layers with magnetoelectric properties. The model comprises the standard Maxwell field coupled to a Chern-Simons field confined to a planar layer. The electromagnetic behavior of the boundary is emulated through the interaction between the Chern-Simons and Maxwell fields, governed by two parameters: the Chern-Simons mass and the coupling constant between the fields. Both parameters can be adjusted to reflect the specific properties of different materials.
We compute the exact propagator of the theory and employ it to investigate several physical properties. Our analysis focuses on phenomena that arise from the presence of external sources coupled to both the Maxwell and Chern-Simons fields, considering various scenarios. In the Chern-Simons sector, the sources emulate defects in the crystal lattice of the material layer.
The main objective of this paper is to present the proposed model and to explore its behavior in the simple context of a single planar material interface. We also suggest possible extensions of the model to more general configurations.
\end{abstract}

\maketitle

\section{\label{I}Introduction}

The use of field-theoretic models, especially those exhibiting gauge symmetry, to describe condensed matter phenomena is an issue well established in the literature. Within this framework, effective field theories with emergent gauge fields have proven to be powerful tools for modeling crystal lattices. Notable examples include the Quantum Hall Effect \cite{ZhangIJMP92,ZeeNPB91,BanerjeeMukherjee1,BanerjeeMukherjee2,PimentelBJP2023} and topological insulators \cite{Review2010,Review2016,ShenLivro,MarinoLivro,PRB2016,PRX2015,PRB2015,PRB2020}. In this context, we highlight the versatile role of Chern-Simons-like models \cite{MCS1,Dune98} in capturing magnetoelectric phenomena \cite{Helder2021,PRXmetamateriais}, among other effects, especially in planar systems.

Chern-Simons-type models are extensively explored in the literature across a wide range of contexts, including the Casimir effect \cite{Casimir2a,Casimir2,Casimir3a,Casimir3,Casimir5,RudneiPRD2012,PLA2000,MarachevskyPRB2019}, quantum electrodynamics \cite{QED3,QED4,QED5,Russo}, the influence of external sources and/or potentials \cite{EPJC802020}, wave propagation phenomena \cite{WegnerPRE2015}, extensions to higher dimensions \cite{CarrollFieldJakiew1,CarrollFieldJakiew2,CarrollFieldJakiew3,BorgesEPJPlus2014,CSKR1,CSKR2,CSKR2a,CSKR3,CSKR4}, and models involving higher-order derivatives \cite{CHderivadasuperior1,CHderivadasuperior2,CHderivadasuperior3,CHderivadasuperior4,CHderivadasuperior5,CHderivadasuperior6}, to name just a few.

Another alternative approach employed in the literature to describe material layers involves models in which fields are coupled to external potentials, particularly those that are spatially localized. Among these, special attention is given to models involving delta-like functions \cite{Milton,Mostepanenko,BordagHeningRobaschik1992,KimballA,Bordag1999,NRVMH,NRVMMH2,Jaffe2004,Caval,Losada2008,CapaFoscoLousada,Fosco2012,GTFABFEB,FoscoRemmaggi,PKKM,DaniloPRD2016,MITescalar}. In the electromagnetic case, such models can be used to simulate the presence of a uniaxial dielectric layer \cite{FABFEB,FABFEB2} or magnetoelectric boundaries \cite{Helder2021,Planar5}, which can also be described by using the so-called theta model \cite{ModeloTheta1,ModeloTheta2,ModeloTheta3,ModeloTheta4}.

In the case of magnetoelectric boundaries, most of the models explored thus far, to the best of the authors' knowledge, introduce a single free parameter that is used to characterize the physical properties of the boundary. Models with more than one free parameter are desirable, as they would offer greater versatility in tuning the electromagnetic properties of a material boundary. Moreover, a theoretical model incorporating two gauge fields may provide a more accurate description of material surfaces, with one of the fields specifically representing the crystalline lattice structure.

In this work, we address this issue in the context of magnetoelectric surfaces. We employ the Maxwell-Chern-Simons planar field, whose dynamics are entirely restricted to the material surface, to modulate its magnetoelectric properties. The choice of a Chern-Simons-like field was motivated by its well-known effectiveness in modeling electromagnetism in planar systems. Additionally, we consider a coupling between the planar field and the standard Maxwell electromagnetic field, which is defined throughout the entire space. This coupling between the two gauge fields is implemented through a Dirac delta-like potential concentrated along the plane where the Chern-Simons field is defined; it exhibits a Chern-Simons-like structure and is governed by a parameter \(\mu\). Thus, the proposed two-gauge-field model introduces two independent free parameters: the Chern-Simons mass \(m\) and the coupling parameter \(\mu\), both of which can be adjusted to fit properties observed in magnetoelectric boundaries.

The propagator of the model is calculated exactly, and it is shown that in the strong coupling limit, the model is equivalent to the electromagnetic field in the presence of a perfectly conducting plate. We use the proposed model to investigate some of its electromagnetic features in the context of various types of defects in the crystal lattice and/or the presence of electric charges. Additionally, we study certain field configurations within this scenario to highlight the magnetoelectric properties of the model.

Usually, models developed to describe planar material media are strictly confined to $(2+1)$ dimensions, as in the study of quantum Hall liquids, topological insulators, graphene under external fields, and superconductivity. In addition, the electromagnetic field is often treated as an externally prescribed, non-dynamical field, with only its projection onto the planar material medium being considered. Representative examples include \cite{Ex1,Ex2,Ex3,Ex4,Ex5,Ex6,Ex7,Ex8,Ex9,Ex10,Ex11}.

The model proposed in this work broadens these possibilities by considering the planar system embedded in $(3+1)$-dimensional space, where the electromagnetic field is defined throughout the entire space and endowed with dynamical terms. This framework enables the investigation of physical phenomena not only on the material surface but also in its surroundings. Within this context, one may, for example, consider external agents relative to the material plane, such as charges and currents, interacting with defects in the crystalline lattice, or even electromagnetic sources located inside the medium. Chern-Simons-like couplings between planar and bulk gauge fields have been employed in the literature to analyze quantum Hall liquids, topological insulators, graphene coupled to external fields, and superconductivity, as previously noted, though in a much more restrictive setting than the one proposed here.

For interested readers, variants of the present model can be explored by modifying the dynamical terms of the planar gauge field, for instance, by considering only a Chern-Simons-type term, as is often done in the description of certain planar magnetoelectric systems. Within the framework proposed here, one may also investigate different planar surfaces and the interactions that arise among them.

We hope that the discussions presented throughout this work will draw attention to the role that two-field models can play in emulating material surfaces with electromagnetic properties across a broader range of contexts.

In Sec. (\ref{II}), we propose the Lagrangian that defines the model and calculate the corresponding propagator exactly. We also consider the dynamical field equations, identifying the polarization and magnetization vectors associated with the model. In Sec. (\ref{III}), we study the interaction between the material surface and stationary charges. Additionally, in section (\ref{IV}), we analyze the interaction between defects on the material surface and charges, as well as the interaction among different defects on the surface itself. In Sec. (\ref{V}), we examine the field solutions induced by a stationary electric charge in the presence of the magnetoelectric material plate. The results highlight some of the magnetoelectric properties that the model can emulate. Sec. (\ref{conclusoes}) is devoted to our final remarks.

In this paper we work in natural units ($\hbar=1$ and $c=1$) in a $3+1$-dimensional Minkowski space-time with metric $\eta^{\rho\nu}=(1,-1,-1,-1)$. The Levi-Civita tensor is denoted by $\epsilon^{\rho\nu\alpha\beta}$ with $\epsilon^{0123}=1$.

\section{\label{II} The modified photon propagator}

In this section, we propose a field-theoretical model that emulates the presence of material layers with magnetoelectric properties. 
We shall consider just a single planar layer and adopt a coordinate system where it lies along the plane $x^{3}=a$.

The electromagnetic gauge field shall be denoted by $A^{\mu}=(A^0,A^1,A^2,A^3)$, and the presence of the material boundary is emulated by a Chern-Simons type potential ${\cal A}^\mu=({\cal A}^0, {\cal A}^1,{\cal A}^2)$, which exhibits gauge symmetry, is defined only along the material layer, where $x^{3}=a$, and couples to the electromagnetic field through a Chern-Simons-like interaction. So, we have a spatial and a planar sector, given by $A^{\mu}$ and ${\cal A}^{\mu}$, respectively. The model is defined by the following Lagrangian density
\begin{eqnarray}
\label{lagrangeana2}
{\cal L}&=& -\frac{1}{4}F^{\mu\nu}F_{\mu\nu}-J_\mu A^\mu-\frac{1}{2\alpha}(\partial_\mu A^\mu)^2+\Bigg[-\frac{1}{4}G^{\mu\nu}G_{\mu\nu}-\mathcal{J}_\mu {\cal A}^\mu-\frac{1}{2\beta}(\partial_\mu {\cal A}^\mu)^2\nonumber\\
&
&+\frac{m}{2}\epsilon^{\mu\nu\alpha 3}{\cal A}_\mu\partial_\nu {\cal A}_\alpha-\frac{\mu}{4}\epsilon^{\mu\nu\alpha 3}A_\mu(\partial_\nu {\cal A}_\alpha)- \frac{\mu}{4}\epsilon^{\mu\nu\alpha 3}{\cal A}_\mu(\partial_\nu A_\alpha)\Bigg]\delta(x^3-a)\ ,
\end{eqnarray}
where $F^{\mu\nu}=\partial^{\mu}A^{\nu}-\partial^{\nu}A^{\mu}$ and $G^{\mu\nu}=\partial^{\mu}{\cal A}^{\nu}-\partial^{\nu}{\cal A}^{\mu}$ represent the field strengths of the fields $A^\mu$ and ${\cal{A}}^\mu$. respectively, $\alpha$ and $\beta$ are gauge-fixing parameters, $J^{\mu}$ is the external source associated with the photon field, $\mathcal{J}^{\mu}$ is the external source associated with the planar sector and the pseudo-scalar constant $\mu$, assumed to be positive, is the coupling constant between the fields that make up the theory. Besides, $m$ is a Chern-Simons-like mass factor for the ${\cal A}^{\mu}$ field. 

The coupling between the photon field and the Chern-Simons field is given by two types of Chern-Simons terms involving both $A^{\mu}$ and ${\cal A}^{\mu}$. These two terms differ from each other by a total derivative, so they contribute equally to the action of the system. We have written them separately for future convenience.

By performing a direct dimensional analysis, taking into account that we are using natural units, and denoting $[\ell]$ as the length dimension, we have
\begin{eqnarray}
[A^{\mu}]=[\ell]^{-1}\ \ ,\ \ [{\cal A}^{\mu}]=[\ell]^{-1/2}\cr\cr
[J^{\mu}]=[\ell]^{-3}\ \ ,\ \ [I^{\mu}]=[\ell]^{-5/2}\cr\cr
[m]=[\ell]^{-1}\ \ ,\ \ [\mu]=[\ell]^{-1/2}\ .
\end{eqnarray}
So, $m$ has a mass dimension indeed, but the coupling constant $\mu$ does not.

From now on, we shall work in the Feynman gauge for both fields, that is, we shall take $\alpha=1$ and $\beta=1$. 

Notice that the Dirac delta function in (\ref{lagrangeana2}) ensures that the terms involving the Chern-Simons field are non-vanishing only along the material boundary, which includes the interaction terms between the Chern-Simons field and the Maxwell field. To understand the role of these delta-like terms in the photon sector, it is convenient to write the dynamical equations associated with both fields as follows
\begin{equation}\label{eqmov}
    \begin{split}
        &\partial_\sigma F^{\sigma\nu}=J^\nu-\frac{\mu}{2}\delta(x^3-a)\tilde{G}^{3 \nu},\\
        &\partial_\sigma G^{\sigma \nu} + m \tilde{G}^{\nu 3} = \mathcal{J}^{\nu}+\frac{\mu}{2}\tilde{F}^{\nu 3 }\Big|_{x^3=a},
    \end{split}
\end{equation}
where $\tilde{G}^{\mu\nu}=\frac{1}{2}\epsilon^{\mu\nu\alpha\beta}G_{\alpha\beta}$ and $\tilde{F}^{\mu\nu}=\frac{1}{2}\epsilon^{\mu\nu\alpha\beta}F_{\alpha\beta}$ stand for the dual field strength tensors of $A^\mu$ and $\mathcal{A}^\mu$, respectively. For the photon sector, the electric and magnetic fields are given by $\mathbf{E} = -\nabla A^{0} - \partial_0 \mathbf{A}$ and $\mathbf{B} = \nabla \times \mathbf{A}$. In the planar sector, the corresponding fields are $\pmb{\mathcal{E}} = -\nabla_\parallel \mathcal{A}^{0} - \partial_0 \pmb{\mathcal{A}}$ and $\mathcal{B} = \partial_{2}\mathcal{A}^1 - \partial_{1}\mathcal{A}^2$. Therefore, equations (\ref{eqmov}) reduce to
\begin{equation}\label{maxeq}
    \begin{split}
        &\nabla.\mathbf{E}(x)=J^0(x) - \frac{\mu}{2}\delta(x^3-a)\mathcal{B}(x_\parallel),\\
        &\nabla\times\mathbf{B}(x)=\mathbf{J}(x)+\partial_0 \mathbf{E}(x)+\frac{\mu}{2}\delta(x^3-a) \;\pmb{\mathcal{E}}(x_\parallel)\times \hat{n},
    \end{split}
\end{equation}
\begin{equation}\label{cherneq}
    \begin{split}
        \hspace{3.1cm}&\nabla_\parallel . \pmb{\mathcal{E}}(x_\parallel)- m \mathcal{B}(x_\parallel)= \mathcal{J}^0(x_\parallel)+\frac{\mu}{2}\;\hat{n}.\mathbf{B}(x)\Big|_{x^3=a},\\
         &(\nabla_{\parallel}\times\hat{n})\mathcal{B}(x_\parallel)=\pmb{\mathcal{J}}(x_\parallel)+\partial_0 \pmb{\mathcal{E}}(x_\parallel)-m(\pmb{\mathcal{E}}(x_\parallel)\times\hat{n})+\frac{\mu}{2}(\mathbf{E}(x)\times\hat{n})\Big|_{x^3=a},
    \end{split}
\end{equation}
where we define the Minkowski vector parallel to the layer, $x_{\parallel}=(x^{0},x^{1},x^{2})$, and $\hat{n}=(0,0,1)$ is the normal space vector perpendicular to the layer.

At this point, some comments are in order. For the photon sector, the presence of the $\delta$-like potential introduces additional terms that depend on the electric and magnetic fields of the planar sector themselves, rather than merely on their derivatives. Conversely, the photon sector contributes to the planar sector through terms that depend on the photon electromagnetic fields themselves, evaluated at the plane $x^{3}=a$. Therefore, from (\ref{cherneq}) and (\ref{maxeq}), the coupling between the photon and Chern-Simons sectors becomes evident.

The model (\ref{lagrangeana2}) exhibits a $\delta$-type divergence in the photon sector due to the planar potential. To understand the role of this divergence in the photon field strength tensor, it is convenient to write the first equation in (\ref{eqmov}) as follows 
\begin{eqnarray}
\label{motionE1}
\partial_{\rho}F^{\rho\nu}=J^\nu+\frac{\mu}{2}\delta\left(x^{3}-a\right)\epsilon^{\nu\rho\alpha 3}\partial_\rho{\cal A}_\alpha \ ,
\end{eqnarray}
which, in terms of electric and magnetic fields, reads
\begin{equation}\label{motionE2}
    \nabla.\;\mathbf{E}(x)=J^0(x)-\frac{\mu}{2}\delta(x^3-a)\big[\nabla_{\parallel}.(\hat{n}\times\pmb{\mathcal{A}}(x_\parallel))\big],
\end{equation}
\begin{equation}\label{motionE3}
    \nabla\times\mathbf{B}(x)=\mathbf{J}(x)+\partial_0 \mathbf{E}(x)+\frac{\mu}{2}\delta(x^3-a)\Big[\partial_0 (\hat{n}\times\pmb{\mathcal{A}}(x_\parallel))+\nabla_{\parallel}\times(\hat{n}\;\mathcal{A}^0(x_\parallel))\Big].
\end{equation}

From equations (\ref{motionE2}) and (\ref{motionE3}), we can identify a polarization $\mathbf{P}$ and magnetization $\mathbf{M}$,
\begin{eqnarray}
\label{Polarization}
\mathbf{P}(x)&=&\frac{\mu}{2}\delta\left(x^{3}-a\right)\left(\hat{n}\times{\pmb{{\cal A}}(x_\parallel)}\right), \  \ \\
\label{Magnetization}
\mathbf{M}(x)&=&\frac{\mu}{2}\delta\left(x^{3}-a\right){\cal A}^0(x_\parallel)\;\hat{n}.
\end{eqnarray}
Thus, rewriting the equations (\ref{motionE2}) and (\ref{motionE3}) in terms of $\mathbf{P}$ and $\mathbf{M}$ we have
\begin{eqnarray}
\label{motionPol}
{{\nabla}}\cdot{\mathbf{E}}(x)&=&J^{0}(x)-{{\nabla_{\parallel}}}\cdot\mathbf{P}(x),\  \ \\
\label{motionMag}
{{\nabla}}\times{\mathbf{B}}(x)&=&{\mathbf{J}}(x)+ \partial_0 \mathbf{E}(x) +\partial_0 \mathbf{P}(x)+\nabla_{\parallel}\times\mathbf{M}(x).
\end{eqnarray}

Thus, the presence of the layer can be interpreted in terms of polarization and magnetization defined along the plane $x^{3}=a$, both of which are given by the Chern-Simons field. We could use equation (\ref{cherneq}) to express $\mathbf{P}(x)$ and $\mathbf{M}(x)$ as functions of $\mathbf{E}(x)$ and $\mathbf{B}(x)$, which would be equivalent to establishing an effective theory for the Maxwell field. However, we will not follow this approach. Instead, we will investigate the theory in its entirety.

To further investigate the theory described by the Lagrangian density (\ref{lagrangeana2}), neglecting surface terms, we will express it in matrix form,
\begin{equation}
\label{model1}
    \mathcal{L}=\frac{1}{2} \mathbb{A}_{\mu}^{t}\mathcal{O}^{\mu\nu}\mathbb{A}_{\nu} - \mathbb{A}_{\mu}^{t} \mathbb{J}^\mu
\end{equation}
where the collunm matrices $\mathbb{A}_\mu$ and $\mathbb{J}^\mu$ are defined by
\begin{equation}
\label{campoecorrente}
    \mathbb{A}_\mu = \begin{bmatrix}
        \mathbb{A}_{_{(1)}\mu}\\
        \mathbb{A}_{_{(2)}\mu}
    \end{bmatrix}
    =\begin{bmatrix}
        A_\mu\\
        \mathcal{A}_\mu
    \end{bmatrix},\;\; \mathbb{J}_\mu = \begin{bmatrix}
        \mathbb{J}_{_{(1)}\mu}\\
        \mathbb{J}_{_{(2)}\mu}
    \end{bmatrix}
    =\begin{bmatrix}
        J_\mu\\
        \delta(x^3-a)\mathcal{J}_\mu
    \end{bmatrix},
\end{equation}
with ${\cal{O}}^{\mu\nu}$ being a $2X2$-square matrix whose elements are differential operators defined by
\begin{eqnarray}
\label{Matrix}
{\mathcal{O}}^{\mu\nu}&=&\left[\begin{array}{rr}
\mathcal{O}^{\mu\nu}_{(11)} & \mathcal{O}_{(12)}^{\mu\nu} \\
\\

\mathcal{O}_{(21)}^{\mu\nu}  & \mathcal{O}_{(22)}^{\mu\nu} 
\end{array}\right]\  \ ,
\end{eqnarray}
such that
\begin{equation}
\label{elementsmatrix}
    \begin{split}
        &\mathcal{O}^{\mu\nu}_{(11)}=\eta^{\mu\nu}\Box,\\
        &\mathcal{O}^{\mu\nu}_{(12)}=\mathcal{O}^{\mu\nu}_{(21)}=-\frac{\mu}{2}\epsilon^{\mu\alpha\nu 3}\delta\left(x^{3}-a\right)\partial_{\|\alpha},\\
        &\mathcal{O}^{\mu\nu}_{(22)}=\delta\left(x^{3}-a\right)(\eta_{\|}^{\ \mu\nu}\Box_{\|} + m\epsilon^{\mu\alpha\nu 3}\partial_{\|\alpha}),
    \end{split}
\end{equation}
where $\eta^{\mu\sigma}_{\|}=\eta^{\mu\sigma}-\eta^{\mu 3}\eta^{\sigma 3}$. The matrix element $\mathcal{O}^{\mu\nu}_{(11)}$ is the operator obtained from the Maxwell Lagrangian with the Feynman gauge $\alpha=1$. On the other hand, $\mathcal{O}^{\mu\nu}_{(12)}$ and $\mathcal{O}^{\mu\nu}_{(21)}$ correspond to the terms resulting from the coupling between the fields $A^\mu$ and ${\cal A}^\mu$. Lastly, the matrix element $\mathcal{O}^{\mu\nu}_{(22)}$ is the one corresponding to the field ${\cal A}^\mu$ with the gauge parameter $\beta=1$.
We note that the derivative in (\ref{elementsmatrix}) is only defined in the Minkowski coordinates parallel to the planar potential.

The propagator of the model is given by the inverse of the operator ${\cal{O}}^{\mu\nu}$ and satisfies the equation
\begin{equation}
\label{propag1}
    \mathcal{O}^{\mu\nu}G_{\nu\sigma}(x,y) = \begin{bmatrix}
    \eta_{\;\sigma}^{\mu}\delta^4(x-y) & 0\\
    \\
    0 & \eta_{\parallel\;\sigma}^{\mu}\delta^3(x_\parallel-y_\parallel)\delta(x^3-a)
    \end{bmatrix}.\;\;\;\;\; 
\end{equation}

For convenience, we split ${\cal{O}}^{\mu\nu}$ into two parts: one corresponding to the non-interacting terms, and the other arising from the interacting terms between the gauge fields,
\begin{eqnarray}
\label{model2}
{\cal{O}}^{\mu\nu}={\cal{O}}^{(0)\mu\nu}+ \Delta{\cal{O}}^{\mu\nu}  ,
\end{eqnarray}
with

\begin{equation}
    \mathcal{O}^{(0)\mu\nu}=\begin{bmatrix}
        \mathcal{O}^{\mu\nu}_{(11)}&0\\
        \\
        0&\mathcal{O}^{\mu\nu}_{(22)}
    \end{bmatrix},\;\;\;\Delta\mathcal{O}^{\mu\nu}=\begin{bmatrix}
        0 & \mathcal{O}^{\mu\nu}_{(12)}\\
				\\
        \mathcal{O}^{\mu\nu}_{(21)}&0
    \end{bmatrix}
\end{equation}

It is straightforward to verify that the operator ${\cal{O}}^{(0)\mu\nu}$ satisfies the differential equation
\begin{equation}
    \mathcal{O}^{(0)\mu\nu}(x)\;G^{(0)}_{\nu\sigma}(x,y)=\begin{bmatrix}
    \eta_{\;\sigma}^{\mu}\delta^4(x-y) & 0\\
    \\
    0 & \eta_{\parallel\;\sigma}^{\mu}\delta^3(x_\parallel-y_\parallel)\delta(x^3-a)
    \end{bmatrix}
\end{equation}
with $G^{(0)}_{\nu\sigma}\left(x,y\right)$ given by 
\begin{equation}\label{mu0}
    G^{(0)}_{\nu\sigma}(x,y)=\begin{bmatrix}
        G_{(M)\nu\sigma}(x,y)&0\\
        0 & G_{(CS)\nu\sigma}(x_\parallel,y_\parallel)
    \end{bmatrix}
\end{equation}
where $G_{(M)\nu\sigma}(x,y)$ is the photon propagator in the Feynman gauge $(\alpha =1)$, and $G_{(CS)\nu\sigma}(x_\parallel,y_\parallel)$ is the Chern-Simons propagator with gauge parameter $\beta =1$. Their Fourier integrals expanded in the coordinates parallel to the plane are given by
\begin{equation}\label{fourierparallel}
    \begin{split}
        &G_{(M)\nu\sigma}(x,y) = \int \frac{d^3 p_\parallel}{(2\pi)^3}\mathcal{G}_{(M)\nu\sigma}(p_\parallel;x^3,y^3) e^{-i p_\parallel(x_\parallel-y_\parallel)},\\
        &G_{(CS)\nu\sigma}(x_\parallel,y_\parallel)=\int \frac{d^3 p_\parallel}{(2\pi)^3}\mathcal{G}_{(CS)\nu\sigma}(p_\parallel) e^{-i p_\parallel(x_\parallel-y_\parallel)},
    \end{split}
\end{equation}
with 
\begin{equation}
\begin{split}
    &\mathcal{G}_{(M)\nu\sigma}(p_\parallel;x^3,y^3)=\frac{e^{-\sqrt{-p_\parallel^2}|x^3-y^3|}}{2\sqrt{-p_\parallel^2}}\;\eta_{\nu\sigma},\\
    &\mathcal{G}_{(CS)\nu\sigma}(p_\parallel)=\frac{-1}{p_\parallel^2-m^2}\Big(\eta_{\parallel\;\nu\sigma}-\frac{m^2}{p_\parallel^2}\frac{p_{\parallel\;\nu}p_{\parallel\;\sigma}}{p_\parallel^2}-\frac{im}{p_\parallel^2}\epsilon_{\nu\sigma\gamma 3}\;p_{\parallel}^{\gamma} \Big).
\end{split}
\end{equation}


Analogously to Eq.~(\ref{fourierparallel}), we express $G_{\nu\sigma}\left(x,y\right)$ and $G^{(0)}_{ \nu\sigma}\left(x,y\right)$ as Fourier integrals in the coordinates parallel to the planar potential.
\begin{eqnarray}
\label{prop2}
G_{\nu\sigma}\left(x,y\right)=\int\frac{d^{3}p_{\parallel}}{\left(2\pi\right)^{3}} \ {\cal{G}}_{\nu\sigma}\left(p_{\parallel};x^{3},y^{3}\right)e^{-ip_{\parallel}\cdot\left(x_{\parallel}-y_{\parallel}\right)} , \\
\label{prop3}
G^{(0)}_{\nu\sigma}\left(x,y\right)=\int\frac{d^{3}p_{\parallel}}{\left(2\pi\right)^{3}} \ {\cal{G}}^{(0)}_{\nu\sigma}\left(p_{\parallel};x^{3},y^{3}\right)e^{-ip_{\parallel}\cdot\left(x_{\parallel}-y_{\parallel}\right)} .
\end{eqnarray}
The functions ${\cal{G}}_{\rho\nu}\left(p_{\parallel};x^{3},y^{3}\right)$ and ${\cal{G}}^{(0)}_{\rho\nu}\left(p_{\parallel};x^{3},y^{3}\right)$ are commonly referred to as reduced propagators \cite{LHCBAFFFAB,Caval}.

By substituting equations (\ref{prop2}) and (\ref{prop3}) into (\ref{propag1}), we obtain 
\begin{equation}
\label{prop5}
    \begin{split}
        \mathcal{G}_{\nu\sigma}(p_\parallel;x^3,y^3)=\mathcal{G}^{(0)}_{\nu\sigma}(p_\parallel;x^3,y^3)- \frac{i\mu}{2}\epsilon^{\beta\alpha\kappa3}p_{\parallel\;\alpha}\;\mathcal{G}_{\nu\beta}(p_{\parallel};x^3,a)\begin{bmatrix}
            0&1\\
            1&0
        \end{bmatrix}
        \mathcal{G}^{(0)}_{\kappa\sigma}(p_\parallel;a,y^3).
    \end{split}
\end{equation}

From Eq. (\ref{prop5}), we obtain the reduced propagator \(\mathcal{G}_{\rho\nu}(p_\parallel; x^3, y^3)\). Evaluating Eq. (\ref{prop5}) at \(y^3 = a\), and applying some algebraic manipulations, we arrive at
\begin{equation}\label{prop6}
    \begin{split}
        \mathcal{G}_{\nu\gamma}(p_\parallel;x^3,a)\;M^{\gamma}_{\;\;\theta}=\mathcal{G}^{0}_{\nu\sigma}(p_\parallel;x^3,a)\eta_{\parallel\;\theta}^{\sigma},
    \end{split}
\end{equation}
where the matrix $M^{\gamma}_{\;\;\theta}$ is defined with respect to the momentum components parallel to the layer
\begin{equation}
\label{defmatr}
    M^{\gamma}_{\;\;\theta}=\begin{bmatrix}
            \eta_{\parallel\;\theta}^{\gamma}&\frac{\mu}{2(p_\parallel^2-m^2)}\big(m\;\eta_{\parallel\;\;\sigma}^\gamma  -\frac{m}{p_\parallel^2}p^{\gamma}_{\parallel}\; p_{\parallel\theta} -i\epsilon^{\gamma}_{\;\theta\alpha 3}p_{\parallel}^{\alpha}\big)\\
            \\
            \frac{i\mu}{4\sqrt{-p_\parallel^2}}\epsilon^{\gamma}_{\;\theta \alpha 3}\;p_{\parallel}^\alpha & \eta_{\parallel\;\theta}^\gamma
    \end{bmatrix}.
\end{equation}

Now we multiply both sides of (\ref{prop6}) by the inverse $(M^{-1})^{\theta}_{\;\;\tau}$ of the matrix (\ref{defmatr}), in the sense that
\begin{equation}
\label{defmatrinv}
    M^\gamma_{\;\;\theta}\;(M^{-1})^{\theta}_{\;\;\tau}=\eta^{\gamma}_{\parallel\;\tau}\begin{bmatrix}
        1&0\\
        0&1
    \end{bmatrix}.
\end{equation}
so, we find
\begin{equation}\label{p1}
    \mathcal{G}_{\nu\beta}(p_\parallel;x^3,a)\;\eta^\beta_{\parallel\;\tau}=\mathcal{G}^{(0)}_{\nu\beta}(p_\parallel;x^3,a)\;\eta^{\sigma}_{\parallel\;\theta}(M^{-1})^\theta_\tau.
\end{equation}

By inserting (\ref{p1}) in eq. (\ref{prop5}), we obtain the following form for the total propagator
\begin{equation}
\label{prop7}
    \mathcal{G}_{\nu\sigma}(p_\parallel;x^3,y^3)=\mathcal{G}^{(0)}_{\nu\sigma}(p_\parallel;x^3,y^3)+\Delta\mathcal{G}_{\nu\sigma}(p_\parallel;x^3,y^3),
\end{equation}
where 
\begin{equation}
\label{defmatr1}
\begin{split}
    \Delta\mathcal{G}_{\nu\sigma}(p_\parallel;x^3,y^3)&=\frac{i\mu}{2}\epsilon^{\tau\rho\alpha 3}p_{\parallel \alpha}\;\eta^{\gamma}_{\parallel\;\theta}\;\mathcal{G}^{(0)}_{\nu\gamma}(p_\parallel;x^3,a)(M^{-1})^{\theta}_{\;\tau}\begin{bmatrix}
        0&1\\
        1&0
    \end{bmatrix}
    \mathcal{G}^{(0)}_{\rho\sigma}(p_\parallel;a,y^3),\\
    &=\begin{bmatrix}
        \Delta\mathcal{G}_{(11)\nu\sigma}(p_\parallel;x^3,y^3)&\Delta\mathcal{G}_{(12)\nu\sigma}(p_\parallel;x^3,y^3)\\
        \Delta\mathcal{G}_{(21)\nu\sigma}(p_\parallel;x^3,y^3)&\Delta\mathcal{G}_{(22)\nu\sigma}(p_\parallel;x^3,y^3)
    \end{bmatrix},
\end{split}
\end{equation}
represents the correction to the free reduced propagator \(\mathcal{G}^{(0)}\) resulting from the interaction between the gauge fields.

By inverting the matrix in Eq. (\ref{defmatr}), substituting the result into Eq. (\ref{defmatr1}), and performing some algebraic manipulations, we obtain its matrix elements as follows,
\begin{equation}
\label{correction 1}
    \begin{split}
      &\Delta \mathcal{G}_{(11)\nu\sigma}(p_\parallel;x^3,y^3)= \frac{-i \chi(p_\parallel)e^{-\sqrt{-p_\parallel^2}(|x^3-a|+|y^3-a|)}}{2\sqrt{-p_\parallel^2}\big\{ 1+p_\parallel^2[2\;\chi(p_\parallel)+\chi(p_\parallel)^2(p_\parallel^2-m^2)]\big\}} \\
      &\times\Big \{i[1+\chi(p_\parallel)(p_\parallel^2-m^2)](p_{\parallel\nu}p_{\parallel\sigma}-p_\parallel^2 \eta_{\parallel\nu\sigma})- m \;\epsilon_{\nu\sigma\alpha 3}p_{\parallel}^\alpha \Big\},
    \end{split}
\end{equation}
\begin{equation}\label{correction 2}
    \begin{split}
        &\Delta \mathcal{G}_{(12)\nu\sigma}(p_\parallel;x^3,y^3)= \frac{i \mu\; e^{-\sqrt{-p_{\parallel}^2}|x^3-a|}}{4 \sqrt{-p_{\parallel}^2}\;(p_\parallel^2-m^2)\big\{ 1+p_\parallel^2[2\;\chi(p_\parallel)+\chi(p_\parallel)^2(p_\parallel^2-m^2)]\big\}}\\
       & \times \Big\{i m \big(\eta_{\parallel \nu \sigma} - \frac{p_{\parallel \nu} p_{\parallel \sigma}}{p_\parallel^2}\big) +[1+\chi(p_\parallel)(p_\parallel^2-m^2)]\epsilon_{\nu\sigma\lambda 3}p_\parallel^\lambda\Big\},
    \end{split}
\end{equation}
\begin{equation}\label{correction 3}
    \begin{split}
        &\Delta \mathcal{G}_{(21)\nu\sigma}(p_\parallel;x^3,y^3)=\frac{i \mu\; e^{-\sqrt{-p_{\parallel}^2}|y^3-a|}}{4 \sqrt{-p_{\parallel}^2}\;(p_\parallel^2-m^2)\big\{ 1+p_\parallel^2[2\;\chi(p_\parallel)+\chi(p_\parallel)^2(p_\parallel^2-m^2)]\big\}}\\
       & \times \Big\{i m \big(\eta_{\parallel \nu \sigma} - \frac{p_{\parallel \nu} p_{\parallel \sigma}}{p_\parallel^2}\big) +[1+\chi(p_\parallel)(p_\parallel^2-m^2)]\epsilon_{\nu\sigma\alpha 3}p_\parallel^\alpha\Big\},
    \end{split}
\end{equation}
\begin{equation}\label{correction 4}
    \begin{split}
        &\Delta \mathcal{G}_{(22)\nu\sigma}(p_\parallel;x^3,y^3)= \frac{-i \chi(p_\parallel)}{(p_\parallel^2-m^2)\big\{ 1+p_\parallel^2[2\;\chi(p_\parallel)+\chi(p_\parallel)^2(p_\parallel^2-m^2)]\big\}}\\
        &\times \Big\{i \big[m^2 + p_\parallel^2(1+\chi(p_\parallel)\big(p_\parallel^2-m^2)\big)\big](\eta_{\parallel\nu\sigma}-\frac{p_{\parallel\nu}p_{\parallel \sigma}}{p_\parallel^2})\\
        &+m\big[2+\chi(p_\parallel)(p_\parallel^2-m^2)\big]\epsilon_{\nu\sigma\lambda 3} \;
        p_\parallel^\lambda\Big\},
    \end{split}
\end{equation}
where we defined the function
\begin{equation}
\label{defChi}
\chi(p_{\|})=\frac{\mu^2}{8\sqrt{-p^2_{\|}}(p^2_{\|}-m^2)}\ .
\end{equation}

Thus, we derived the propagator of the theory, \( G_{\nu\sigma}(x,y) \), which is given by
\begin{eqnarray}
\label{greencompleto}
G_{\nu\sigma}\left(x,y\right)=\int\frac{d^3p_{\|}}{{(2\pi)}^3}\Big({\cal{G}}^{(0)}_{\nu\sigma}(p_{\|};x^3,y^3)+\Delta{\cal{G}}_{\nu\sigma}(p_{\parallel};x^{3},y^{3})  \Big)e^{-i p_\parallel(x_\parallel-y_\parallel)},
\end{eqnarray}
with \({\cal{G}}^{(0)}_{\ \nu\sigma}(p_{\|};x^3,y^3)\) given by equation (\ref{mu0}), and the matrix elements of \(\Delta{\cal{G}}_{\nu\sigma}(p_{\|};x^3,y^3)\) given by (\ref{correction 1}), (\ref{correction 2}), (\ref{correction 3}), and (\ref{correction 4}).

The equation (\ref{greencompleto}) represents the exact propagator of the theory. It consists of the propagator \(\mathcal{G}^{(0)}_{\nu\sigma}\), which encodes the dynamics of the non-interacting gauge fields, and a correction term \(\Delta\mathcal{G}_{\nu\sigma}\), which accounts for the interaction between fields induced by Chern-Simons-like terms in the lagrangian (\ref{lagrangeana2}).

Usually, models involving delta-like couplings in quadratic terms of the fields are known to correspond to specific types of boundary conditions when certain limits of the coupling parameter are taken \cite{GTFABFEB,FABFEB,Caval,FABFEB2,FABFEB}. The present case is not different. Setting \(\mu = 0\) in the propagator causes the term \(\Delta{\cal{G}}_{\nu\sigma}(p_{\parallel};x^{3},y^{3})\) to vanish entirely, leaving only \({\cal{G}}^{(0)}_{\nu\sigma}(p_{\|};x^3,y^3)\), which encodes the information of the dynamics of two non-interacting gauge fields. Conversely, in the limit \(\mu \to \infty\), the anti-diagonal elements of (\ref{defmatr1}) vanish, and the propagator (\ref{greencompleto}) in the electromagnetic sector becomes the usual Maxwell propagator in the Feynman gauge corrected by the presence of a perfect conducting plane in $x^3=a$ \cite{FABFEB}
\begin{equation}\label{limit}
    \begin{split}
        \lim_{\mu\to\infty} G_{(11)\nu\sigma}(x,y)&=\int \frac{d^3p_\parallel}{(2\pi)^3}\Bigg[\frac{e^{-\sqrt{-p_\parallel^2}|x^3-y^3|}}{2\sqrt{-p_\parallel^2}}\; \eta_{\nu\sigma}\\
        &+ \frac{e^{-\sqrt{-p_\parallel^2}(|x^3-a|+|y^3-a|)}}{2\sqrt{-p_\parallel^2}}\Big(\frac{p_{\parallel\nu}p_{\parallel\sigma}}{p^2_\parallel}-\eta_{\parallel\nu}^{\sigma}\Big) \Bigg] e^{-ip_\parallel(x_\parallel-y_\parallel)}.
    \end{split}
\end{equation}

On the other hand, from the perspective of the planar gauge sector, its physical descriptive power is lost. In this limit, the dynamical terms of the planar sector reduce to a pure gauge contribution, yielding no physical observables. The coupling parameter \( \mu \) modifies the mass of the planar gauge sector such that, as \( \mu \to \infty \), the field acquires infinite inertia, thereby preventing its propagation,
\begin{equation}\label{pure gauge}
    \lim_{\mu\to\infty}G_{(22)\nu\sigma}(x,y)=-\int \frac{d^3p_\parallel}{(2\pi)^3} \frac{p_{\parallel\;\nu}}{p_\parallel^2}\frac{p_{\parallel\;\sigma}}{p_\parallel^2}\;e^{-ip_\parallel(x_\parallel-y_\parallel)}.
\end{equation}

This same concept applies in the context of Maxwell-Chern-Simons electrodynamics, where the Chern-Simons mass parameter \( m \) is assumed to be sufficiently large to cease the propagation of the field
\begin{equation}
    \lim_{m\to\infty}\mathcal{G}_{(CS)\nu\sigma}(p_\parallel)=-\frac{p_{\parallel\;\nu}}{p_\parallel^2}\frac{p_{\parallel\;\sigma}}{p_\parallel^2}.
\end{equation}

Thus, the presence of the planar gauge field and its interaction with the electromagnetic field provide a field-theoretical description of a perfectly conducting plate in the strong coupling regime, suppressing magnetoelectric effects.

With the complete propagator (\ref{greencompleto}), one can study a wide range of physical phenomena in the presence of a material plate modeled by the gauge field couplings explored in this paper. From now on, we will focus on the interactions between the field sources in this underlying scenario. The interpretations for the Maxwell sources are straightforward, while the sources for the Chern-Simons field could be used to emulate defects localized along the material plate.

\section{\label{III} Interaction between a point-like charge the potential}

In this section, we consider the interaction energy between a point-like stationary electric charge and the planar potential. We begin by noting that the contribution to the system's energy due to the presence of an external current is given by \cite{LHCBFABplate,LHCBFABplate2,FABAAN1,FABAAN2,LHCFABHel,LHCFABAFFEpjc,LHCFABBjp,FABGFHPrd}
\begin{equation}\label{energy}
    E_0 =-\frac{1}{2T}\int d^4x\;d^4y \;(\mathbb{J}^\rho(x))^t\; G_{\rho\nu}(x,y) \mathbb{J}^\nu(y)
\end{equation}
where \(T\) is the time variable and the limit \(T \to \infty\) is implicitly assumed, \(G_{\rho\nu}(x,y)\) is the Green's function of the theory described in (\ref{greencompleto}), and \(\mathbb{J}^{\rho}(x)\) is the external source defined in (\ref{campoecorrente}), which accounts for both the photonic and planar current densities.

From this point on, we shall investigate the interaction between a point-like source in the photon sector and the $\delta$-like Chern-Simons field. To this end, we consider a system without sources in the Chern-Simons sector and a stationary point charge in the photon sector, located at position $\mathbf{b}$. Accordingly, we set $\mathcal{J}^\mu(x)=0$ and $J^\mu(x)=Q\eta^{\mu}_{\ 0}\delta^3(\mathbf{x}-\mathbf{b})$, where $Q$ is the electric charge magnitude. Thus, we obtain
\begin{equation}
\label{source1}
\mathbb{J}^\mu=\begin{bmatrix}
    Q \eta^{\mu}_{\;0}\;\delta^3(\mathbf{x}-\mathbf{b})\\
    \\
    0
\end{bmatrix}.
\end{equation}
Substituting Eqs. (\ref{source1}) and (\ref{defmatr1}) into (\ref{energy}), discarding the self-energy term for the point charge and performing the integrals in $d^{3}{\mathbf x}$, $d^{3}{\mathbf y}$, $dx^{0}$, $dp^{0}$, $dy^{0}$, we obtain 
\begin{equation}
\label{energyC}
    E_{Q\mu}=-\frac{Q^2\mu^2}{4} \int \frac{d^2 \mathbf{p}_\parallel}{(2\pi)^2} \frac{(8|\mathbf{p}_\parallel| +\mu^2)\;e^{-2|\mathbf{p}_\parallel|R_{\perp}}}{|\mathbf{p}_\parallel|\big[64(\mathbf{p}_\parallel^2+m^2)+16\mu^2 |\mathbf{p}_\parallel|+\mu^4\big]},
\end{equation}
where ${R_{\perp}}=\mid b^3-a\mid$ represents the distance between the planar potential and the charge, and the subscript $Q\mu$ denotes the interaction energy between the planar potential and the charge. Using polar coordinates and performing the relevant integrations, we obtain the following expression for the interaction energy between the external source and the planar potential
\begin{eqnarray}
\label{energyC2}
E_{Q\mu}(Q,\mu,m,{R_{\perp}})=-\frac{Q^2 \mu^2 e^{\frac{1}{4} {R_{\perp}}\mu ^2}}{64 \pi}Re\Big\{e^{2imR_{\perp}}\Gamma\big[0,\frac{R_{\perp}}{4}(\mu^2+8im)\big] \Big\}
\end{eqnarray}
with $Re$ standing for the real part, and the function $\Gamma\left[a,z\right]$ representing the generalized incomplete Gamma function \cite{ArfkenLivro,GradLivro}, which is defined by 
\begin{eqnarray}
\Gamma\left[a,z\right]=\int^{\infty}_{z} dt\;\frac{e^{-t}}{t^{1-a}}.
\end{eqnarray}

Eq. (\ref{energyC2}) provides the exact result for the interaction energy between a point-like charge and the planar gauge field \(\mathcal{A}_\mu\) described by the model (\ref{lagrangeana2}). The corresponding force between the point-like charge and the planar potential is given by
\begin{equation}\label{forcem}
    \begin{split}
        F_{Q\mu}(Q,&\mu,m,R_{\perp})=-\frac{\partial}{\partial R_{\perp}}E_{Q\mu}(Q,\mu,m,R_{\perp}),\\
        &=-\frac{Q^2\mu^2}{64\pi R_{\perp}}+\frac{Q^2\mu^2}{256\pi} Re\Bigg\{ \left(\mu^2 +8 i m\right)e^{\frac{1}{4} {R_{\perp}} \left(\mu^2 +8 i m\right)} \Gamma\bigg[0,\frac{1}{4} {R_{\perp}} \left(\mu^2 +8 i m\right)\bigg]\Bigg\}.
    \end{split}
\end{equation}


Let us now provide some remarks regarding the energy and force discussed above. The first term in Eq. (\ref{forcem}) is inversely proportional to the distance, indicating that it decays more slowly than the Coulomb force. The second term, on the other hand, further suppresses the interaction while preserving its attractive nature and ensuring that the decay with distance is equal to or faster than the Coulomb behavior. Both the Chern-Simons mass $m$ and the coupling constant $\mu$ act as inertia parameters: as they increase, the range of the interaction becomes shorter.

\begin{figure}[H]
	\centering
		\includegraphics[width=0.5\textwidth]{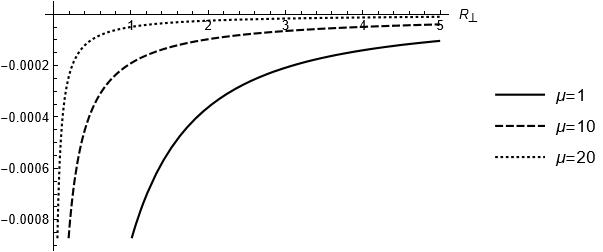}
	\caption{Energy (\ref{energyC2}), normalized by $Q^2\mu^2$ as a function of $R_\perp$, for $m=1$ with $\mu=1$ (solid line), $\mu=10$ (dashed line) and $\mu=20$ (dotted line).}
	\label{fig:Energia_Interacao}
\end{figure}

\begin{figure}[H]
	\centering
		\includegraphics[width=0.5\textwidth]{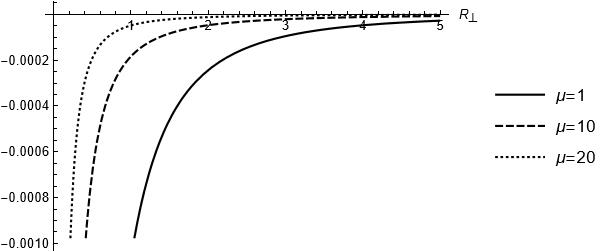}
	\caption{Force (\ref{forcem}), normalized by $Q^2\mu^2$ as a function of $R_\perp$, for $m=1$ with $\mu=1$ (solid line), $\mu=10$ (dashed line) and $\mu=20$ (dotted line).}
	\label{grafico1}
\end{figure}

In Figures (\ref{fig:Energia_Interacao}) and (\ref{grafico1}), we plot the interaction energy (\ref{energyC2}) and the corresponding force (\ref{forcem}), respectively, both normalized by \( Q^2\mu^2 \), as functions of \( R_\perp \), with \( m = 1 \) and for several values of the coupling parameter \( \mu \). It can be observed that the interaction is always attractive.

As discussed earlier, the coupling parameter $\mu$ modifies the mass of the planar gauge field, effectively contributing to its inertia, particularly in the strong coupling limit (\(\mu \to \infty\)). In this regime, the magnetoelectric properties of the theory are suppressed, and the model becomes equivalent to a photon field interacting with a perfectly conducting plate. Consequently, the interaction energy in Eq. (\ref{energyC2}) and the force in Eq. (\ref{forcem}) reduce to the standard Coulomb interaction between a stationary point charge and a perfectly conducting plate, independently of the Chern-Simons mass \(m\), 
\begin{equation}\label{Emuinf}
    E_{Q\mu}(Q,\mu=\infty,m,R_\perp)=-\frac{Q^2}{16\pi R_\perp},
\end{equation}
\begin{equation}\label{Fmuinf}
    F_{Q\mu}(Q,\mu=\infty,m,R_\perp)=-\frac{Q^2}{16\pi R_{\perp}^2}.
\end{equation}

On the other hand, when we consider \(\mu \to 0\), there is no coupling present between the photon and the planar gauge fields. As a result, the theory exhibits no interaction between the charge and the planar potential, and the propagator given in Eq. (\ref{greencompleto}) reduces to Eq. (\ref{mu0}).

For \( m = 0 \), the energy and force, (\ref{energyC2}) and (\ref{forcem}), are given by
\begin{equation}\label{emo}
    E_{Q\mu}(Q,\mu,m=0,{R_{\perp}})=-\frac{Q^2\mu^2}{64\pi}e^{\frac{{R_{\perp}}\mu^2}{4}}  \Gamma\left[0,\frac{{R_{\perp}}\mu^2}{4} \right],
\end{equation}
\begin{equation}\label{fmo}
    F_{Q\mu}(Q,\mu,m=0,{R_{\perp}})=-\frac{Q^2 \mu ^2}{64 \pi  {R_{\perp}}}+\frac{Q^2 \mu ^4}{256 \pi } e^{\frac{{R_{\perp}} \mu ^2}{4}}\Gamma\left[0,\frac{{R_{\perp}} \mu ^2}{4}\right].
\end{equation}

The equations (\ref{emo}) and (\ref{fmo}) recover the Coulomb interaction in (\ref{Emuinf}) and (\ref{Fmuinf}), respectively, in the limit \(\mu \to \infty\). 

\begin{figure}[H]
	\centering
		\includegraphics[width=0.5\textwidth]{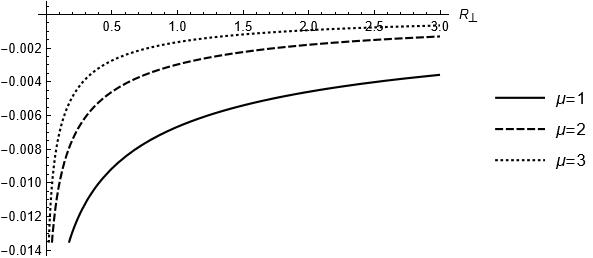}
	\caption{Energy (\ref{emo}), normalized by $Q^2\mu^2$, as a function of $R_\perp$ for $\mu=1$ (solid line), $\mu=2$ (dashed line) and $\mu=3$ (dotted line).}
	\label{graficoEmo}
	\end{figure}

\begin{figure}[H]
	\centering
        \includegraphics[width=0.5\textwidth]{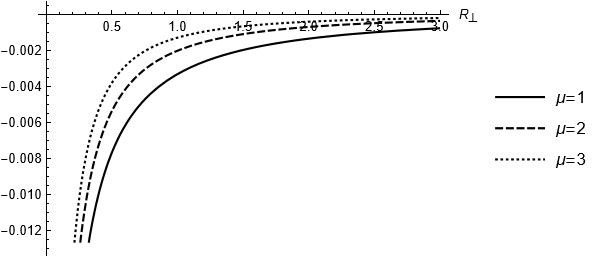}
	\caption{Force (\ref{fmo}), normalized by $Q^2\mu^2$, as a function of $R_\perp$ for $\mu=1$ (solid line), $\mu=2$ (dashed line) and $\mu=3$ (dotted line).}
	\label{graficoFmo}
\end{figure}

In Figures \ref{graficoEmo} and \ref{graficoFmo}, we plot the energy (\ref{emo}) and the force (\ref{fmo}), respectively, both normalized by $Q^2\mu^2$, as functions of \(R_\perp\) for some values of \(\mu\).

\section{Source-Source Interaction}
\label{IV}

In this section, we shall investigate the interaction between two stationary sources in the presence of the planar potential.

\subsection{Two Point-Like charges in the Chern-Simons Sector}

As a first case, we investigate the interaction energy between two stationary point charges in the planar gauge sector. This configuration is obtained by setting \(J^\mu = 0\) and \(\mathcal{J}^\mu = \sum_{i=1}^2 q_i \eta^\mu_{\parallel 0} \delta^2(\mathbf{x}_\parallel - \mathbf{b}_{\parallel i})\), where the source strengths \( q_1 \) and \( q_2 \) have dimension \( [\ell]^{-1/2} \). Thus, the 2-column matrix current for the present case is
\begin{equation}
\label{source2}
    \mathbb{J}^\mu=\begin{bmatrix}
        0\\
        \\
        \sum_{i=1}^2 q_i\eta^\mu_{\parallel\;0} \delta^2(\mathbf{x}_\parallel-\mathbf{b}_{\parallel{i}})\delta(x^3-a)
    \end{bmatrix}.
\end{equation}

By inserting Eqs. (\ref{source2}) and (\ref{greencompleto}) into (\ref{energy}), discarding the self-energies of each point charge and performing the integrals in $d^{3}{\mathbf x}$, $d^{3}{\mathbf y}$, $dx^{0}$, $dp^{0}$, $dy^{0}$, we obtain the following expression for the interaction energy
\begin{equation}
\label{eq1q2}
    E_{q_{1}q_{2}}=\frac{2 q_1 q_2}{\pi^2}\int d^2\mathbf{p}_\parallel \;\frac{(8|\mathbf{p}_\parallel|+\mu^2)e^{i \mathbf{p}_\parallel.\mathbf{R}_\parallel}}{|\mathbf{p}_\parallel|\big[(8|\mathbf{p}_\parallel|+\mu^2)^2+64m^2 \big]},
\end{equation}
where \(\mathbf{R}_\parallel = \mathbf{b}_{\parallel 1} - \mathbf{b}_{\parallel 2}\) is the in-plane distance vector between the charges, with the subscript \(q_1 q_2\) indicating the interaction energy between the two point-like charges in the planar gauge sector. To evaluate Eq. (\ref{eq1q2}), we adopt a coordinate system in which the angle between \(\mathbf{R}_\parallel\) and \(\mathbf{p}_\parallel\) is denoted by \(\phi\). This convention will be used throughout the remainder of the paper. By switching to polar coordinates and performing the integration over both the angular and radial variables in Eq. (\ref{eq1q2}), we obtain the following expression
\begin{equation}
\label{energypointchern2}
    E_{q_{1}q_{2}}(m,\mu,R_\parallel)=\frac{q_1 q_2}{4}Re\Bigg\{H_0\Big[(\frac{\mu^2}{8}+ i m )R_\parallel\Big] -Y_0\Big[(\frac{\mu^2}{8}+ i m )R_\parallel\Big] \Bigg\},
\end{equation}
where $H_\nu$ and $Y_\nu$  denote the Struve and Neumann functions, respectively \cite{ArfkenLivro,GradLivro}, respectively, and $R_\parallel=|\mathbf{R}_\parallel|$ is the distance between sources.

Now, we make some considerations regarding the result in Eq. (\ref{energypointchern2}). Since the arguments of the Struve and Neumann functions are complex, they introduce an exponential decay. The interaction energy diverges at short distances, \( R_\parallel \), and gradually vanishes as the distance increases. The nature of the interaction depends on the signs of \( q_1 \) and \( q_2 \). For the present system, it follows the same pattern as in Maxwell electrodynamics: like-sign sources \(q_1 q_2 >0\) experience a repulsive interaction, whereas opposite-sign sources \(q_1 q_2 >0\) experience an attractive one.

The sources considered here are related to the planar gauge sector, where the coupling parameter \(\mu\) plays a pivotal role, particularly in the strong coupling limit. As previously discussed, the strength of the coupling between the two gauge fields, governed by \(\mu\), directly influences the propagator of the planar field. In the strong coupling regime (\(\mu \to \infty\)), the propagation of the planar gauge field is entirely suppressed, regardless of other system parameters such as the Chern-Simons mass $m$. As a result, the interaction energy given in (\ref{energypointchern2}) vanishes in the limit \(\mu\to\infty\), independently of $m$. 

In contrast, under the decoupling condition (\(\mu = 0\)), the interaction energy in (\ref{energypointchern2}) reduces to well-known result for the interaction between stationary point-like sources in Maxwell-Chern-Simons electrodynamics, as established in the literature \cite{EPJC802020}. This expression is readily obtained by taking \(\mu = 0\) in (\ref{energypointchern2})
\begin{equation}
\label{classmaxchernint}
    E_{q_1 q_2}(m,\mu=0,R)=\frac{q_1 q_2}{2\pi} K_0(m R_\parallel),
\end{equation}
where $K_0(x)$ denotes the modified Bessel function of the second kind of order zero \cite{ArfkenLivro,GradLivro}.

The massless case with no field coupling ($\mu=0$), must be treated with care due to the emergence of logarithmic divergences. By employing the expansion $K_{0}(x)\approx\ln(2)-\ln(x)-\gamma+{\cal O}(x^2)$ in (\ref{classmaxchernint}) we can write
\begin{eqnarray}
E_{q_1 q_2}(m,\mu=0,R)\approx\frac{q_1 q_2}{2\pi} [\ln(2)-\ln(m R_\parallel)-\gamma+\ln(ma)]\cr\cr
\approx-\frac{q_1 q_2}{2\pi}\ln\bigg(\frac{R_\parallel}{a}\bigg)
\end{eqnarray}
where in the first line we added a constant (independent of $R_\parallel$) and $a$ is an arbitrary constant with dimension of length, which does not contribute to the force between the Chern-Simons sources and $\gamma$ is the Euler constant.

\begin{figure}[H]
    \centering
    \includegraphics[width=0.5\linewidth]{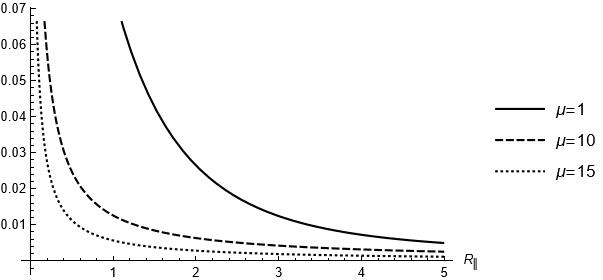}
    \caption{Energy (\ref{energypointchern2}), normalized by $q_1 q_2$, as a function of $R_\parallel$ with $m=1$ and $\mu=1$ (solid line), $\mu=10$ (dashed line) and $\mu=15$ (dotted line).}
    \label{pontochern3d}
\end{figure}

Figure \ref{pontochern3d} displays the interaction energy given by Eq. (\ref{energypointchern2}), normalized by $q_1 q_2$, as a function of $R_\parallel$.

AQUI

\subsection{Point-Like Sources in Distinct Sectors}

Moving forward, we examine the interaction between a point-like charge in the photon sector, located at \( \mathbf{x} = \mathbf{b}_1 \), as previously studied in the context of the plane-charge interaction, and a point-like charge in the Chern-Simons sector, situated at \( \mathbf{x}_\parallel = \mathbf{b}_{\parallel 2} \), as analyzed above. Accordingly, the source terms are given by \( J^\mu(x) = Q\, \eta^\mu_{\;0}\, \delta^3(\mathbf{x} - \mathbf{b}_1) \) and \( \mathcal{J}^\mu (x) = q\, \eta^{\mu}_{\;\parallel 0}\, \delta^2(\mathbf{x}_\parallel - \mathbf{b}_{\parallel 2}) \). In terms of the 2-component current matrix, this reads
\begin{equation}
\label{source3}
    \mathbb{J}^\mu=\begin{bmatrix}
        Q \eta^\mu_{\;0}\delta^3(\mathbf{x}-\mathbf{b}_1)\\
        \\
         q\eta^\mu_{\parallel\;0} \delta^2(\mathbf{x}_\parallel-\mathbf{b}_{\parallel{2}})\delta(x^3-a)
    \end{bmatrix}.
\end{equation}

Here, we shall compute only the contribution to the energy that involves both sources. To obtain the total energy of the system, this term must be supplemented by the plate-charge interaction energy calculated in Section \ref{III}, which contributes solely to the force acting on the electric charge. The Chern-Simons electric charge is confined to the plane $x^{3}=a$ and is subject only to the force that will be analyzed in this section.   

By inserting Eqs. (\ref{source3}) and (\ref{greencompleto}) in (\ref{energy}), discarding all terms that do not involve both sources (i.e., those that do not depend simultaneously on $Q$ and $q$) and performing the integrals over $d^{3}{\mathbf x}$, $d^{3}{\mathbf y}$, $dx^{0}$, $dp^{0}$, $dy^{0}$, $d^2\mathbf{p}_\parallel$, we arrive at the following expression for the interaction energy
\begin{equation}
\label{energyQqfinal}
    E_{Qq}(m,\mu,R_\perp,R_\parallel )=\frac{qQm\mu}{16\pi}\int_{0}^{\infty}dp_\parallel\;\frac{e^{-p_\parallel R_{\perp}}J_{0}(R_{\parallel} p_{\parallel})}{(p_\parallel+\frac{\mu^2}{8})^2+m^2}.
\end{equation}
where \( R_\parallel = |\mathbf{b}_{\parallel 1} - \mathbf{b}_{\parallel 2}| \) denotes the in-plane parallel distance between the two charges, and \( R_\perp = |b^3 - a| \) corresponds to the transverse distance between the point-like charge in the photon sector and the planar potential, as defined in (\ref{energyC}). The subscript \( Qq \) refers to the interaction between the two point-like sources introduced in (\ref{source3}).

From this point onward, we highlight some features of the interaction energy given in Eq.~(\ref{energyQqfinal}). The expression is free of poles and depends not only on the parameters $\mu$ and $m$  in the denominator of the integrand, but also exhibits a linear dependence on these parameters. Although the Chern-Simons mass \(m\) appears only through the dynamical equation of the planar gauge field and does not appear explicitly in the equations governing the photon field, it still affects the interaction energy between the two sources across the two field sectors, effectively acting as a coupling parameter. The interaction vanishes both in the massless limit $(m=0)$ and in the strong mass limit \(m \to \infty\), the latter being expected due to the suppression of planar gauge field propagation caused by its large mass.

Particular attention should be given to the case where the charges are superposed. By setting \( R_\perp = R_\parallel = 0 \)  in Eq. (\ref{energyQqfinal}) and performing the integration over \( p_\parallel \) yields the result
\begin{equation}\label{orig}
    E_{Qq}(m,\mu,R_\perp=0,R_\parallel=0)=\frac{qQ\mu}{16\pi}\arctan{\Big(\frac{8m}{\mu^2}\Big)}.
\end{equation}

From (\ref{orig}), the interaction remains finite at all distances, including in the overlapping limit, where divergences typically arise. This finiteness is ensured by the presence of the coupling \( \mu \) and the Chern-Simons mass \( m \), which together regularize the interaction in (\ref{energyQqfinal}). In the overlapping configuration, as expected, the interaction vanishes both in the decoupling limit of the gauge fields (\( \mu = 0 \)) and in the strong coupling regime (\( \mu \to \infty \)), where gauge field propagation is suppressed and the system behaves like a perfectly conducting plate. However, due to the asymptotic behavior of the \( \arctan \) function, the interaction does not vanish in the strongly massive scenario. Instead, it goes linearly with the coupling parameter $\mu$.

\begin{figure}[H]
    \centering
    \includegraphics[width=0.5\linewidth]{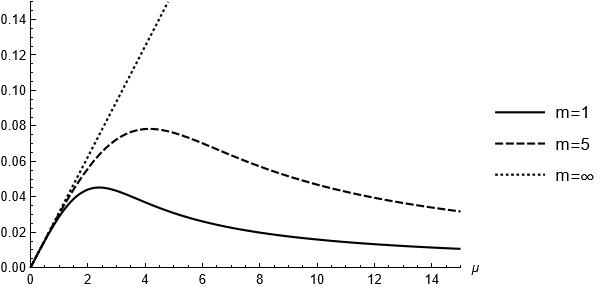}
    \caption{Energy (\ref{orig}), normalized by $qQ$, as a function of $\mu$ for $m=1$ (solid line), $m=5$ (dashed line) and $m=\infty$ (dotted line).}
    \label{plotQqmu}
\end{figure}

\begin{figure}[H]
    \centering
    \includegraphics[width=0.5\linewidth]{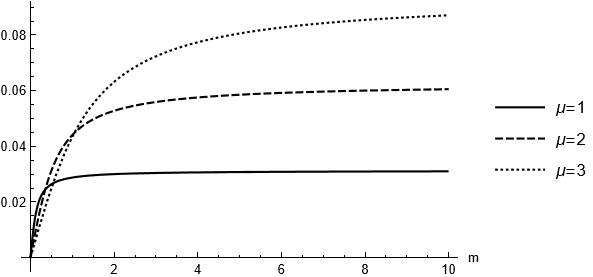}
    \caption{Energy (\ref{orig}), normalized by $qQ$, as a function of $m$ for $\mu=1$ (solid line), $\mu=2$ (dashed line) and $\mu=3$ (dotted line).}
    \label{PlotQqm}
\end{figure}

To illustrate the discussion above, Figures \ref{plotQqmu} and \ref{PlotQqm} display the plot of equation (\ref{orig}), normalized by $Qq$, as a function of $\mu$ and $m$, respectively.

\begin{figure}[H]
    \centering
    \includegraphics[width=0.5\linewidth]{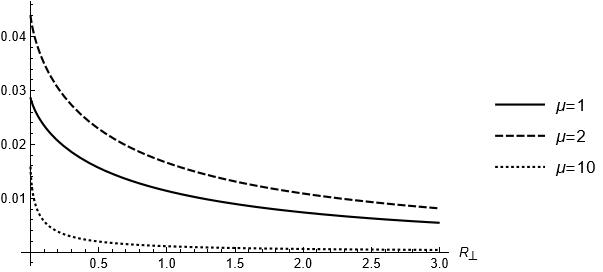}
    \caption{Plot of (\ref{energyQqfinal}), normalized by $qQ$, as a function of $R_\perp$ for $R_\parallel=0$ and $m=1$, with $\mu=1$ (solid line), $\mu=2$ (dashed line) and $\mu=10$ (dotted line).}
    \label{peaksmu}
\end{figure}

\begin{figure}[H]
    \centering
    \includegraphics[width=0.5\linewidth]{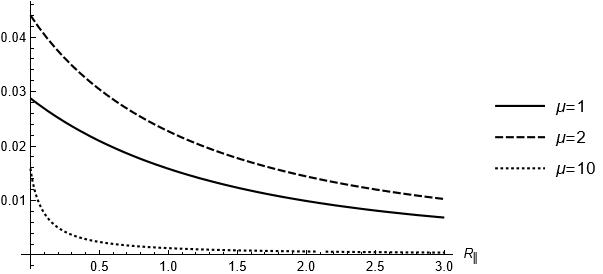}
    \caption{Plot of (\ref{energyQqfinal}), normalized by $qQ$, as a function of $R_\parallel$ for $R_\perp=0$ and $m=1$, with $\mu=1$ (solid line), $\mu=2$ (dashed line) and $\mu=10$ (dotted line).}
    \label{peaksm}
\end{figure}

The energy expression in Eq. (\ref{energyQqfinal}) depends differently on two distance parameters: \( R_{\parallel} \), which represents the in-plane separation between the sources, and \( R_\perp \), which denotes the perpendicular distance between the Maxwell field source and the planar gauge field source. Figures \ref{peaksmu} and \ref{peaksm} display plots of the interaction energy given by Eq.(\ref{energyQqfinal}), normalized by \( qQ \), as functions of \( R_\perp \) or \( R_\parallel \), respectively. In both cases, we set \( m = 1 \) and fix the other distance parameter to zero, exploring the behavior of the energy for different values of \( \mu \). These plots illustrate the behavior of the perpendicular and parallel components of the force relative to the planar gauge field.

AQUI

\subsection{Interaction Between Two Topological Sources}

Here, we consider the interaction energy between two non-trivial point-like sources, which we refer to as topological sources. We begin by analyzing a system composed of two such sources, both associated with the planar sector and located at positions \(\mathbf{b}_{\parallel 1}\) and \(\mathbf{b}_{\parallel 2}\), as follows,
\begin{equation}\label{nontrivial}
    \begin{split}
        &\mathcal{J}^\mu(x_\parallel) = \sum_{i=1}^{2}\epsilon^{\mu\alpha\beta 3}V_{\alpha}^{(i)}\partial_{\parallel\beta}\delta^2(\mathbf{x}_\parallel-\mathbf{b}_{\parallel i}),\\
        &J^\mu(x)=0,
    \end{split}
\end{equation}
where \(V_\alpha^{(1)} = (V^{(1)0}, -\mathbf{V}^{(1)})\) and \(V_\alpha^{(2)} = (V^{(2)0}, -\mathbf{V}^{(2)})\) are constant Minkowski 3-pseudovectors with dimension $[\ell]^{1/2}$. As discussed in refrence \cite{EPJC802020}, the sources in (\ref{nontrivial}) can be obtained through dimensional reduction from the point-like source proposed in \cite{CSKR3} for the Kalb-Ramond field. 

Due to the antisymmetry of the differential operator $\epsilon^{\mu\alpha\beta 3}\partial_{\parallel\mu}\partial_{\parallel\beta}$, it follows that $\partial^\mu \mathcal{J}_\mu(x_\parallel) = 0$, so the source (\ref{nontrivial}) satisfies the continuity equation.  From the perspective of the planar gauge field, this implies that (\ref{nontrivial}) gives rise to an intrinsically conserved quantity, $\int d^2\mathbf{x}\; \mathcal{J}^0(x_\parallel)$. Moreover, in a $(2+1)$-dimensional spacetime, when the coupling term $\mathcal{J}^\mu \mathcal{A}_\mu$ in curved space, by replacing $\eta$ with a general metric $g$, we observe that it does not couple to the gravitational field. This behavior is analogous to that of the Chern-Simons term \cite{ZEE}. For this reason, we refer to (\ref{nontrivial}) as a topological source.

Considering (\ref{nontrivial}), the four-component current matrix is given by
\begin{equation}\label{sourceT}
    \mathbb{J}^\mu(x)=\begin{bmatrix}
       0\\
       \\
       \sum_{i=1}^{2}\epsilon^{\mu\alpha\beta 3}V_{\alpha}^{(i)}\partial_{\parallel\beta}\delta^2(\mathbf{x}_\parallel-\mathbf{b}_{\parallel i})\delta(x^3-a)
    \end{bmatrix}.
\end{equation}

By inserting Eqs. (\ref{sourceT}) and (\ref{greencompleto}) into (\ref{energy}), discarding the self-energy term for each charge and performing the integrals over $d^{3}{\mathbf x}$, $d^{3}{\mathbf y}$, $dx^{0}$, $dp^{0}$, $dy^{0}$ and $d^2\mathbf{p}_\parallel$, we obtain the following expression for the interaction energy
\begin{equation}\label{etoptop}
    \begin{split}
        &E_{V^{(1)}V^{(2)}}(m,\mu,R_\parallel)=\frac{V^{(1)0}V^{(2)0}}{4}Re\Big\{\frac{2}{R_\parallel}(\mu^2/8+i m)H_{-1}((\mu^2/8+i m)R_\parallel)\\
        &+(\mu^2/8+i m)^2\Big[Y_0((\mu^2/8+i m)R_\parallel)+H_{-2}((\mu^2/8+i m)R_\parallel)\Big] \Big\}\\
        &+\frac{\mathbf{V}^{(1)}.\mathbf{V}^{(2)}}{4R_\parallel}Re\Big\{(\mu^2/8+i m)\Big[2\Big(H_{-1}((\mu^2/8+i m)R_\parallel)+2Y_{1}((\mu^2/8+i m)R_\parallel)\Big)\\
        &-(\mu^2/8+i m)R_\parallel\Big(H_{-2}((\mu^2/8+i m)R_\parallel)+Y_{0}((\mu^2/8+i m)R_\parallel)\Big) \Big]\Big\}\\
        &+\frac{\Big(\mathbf{V}^{(1)}.\mathbf{R}_\parallel\Big)\Big(\mathbf{V}^{(2)}.\mathbf{R}_\parallel\Big)}{R^2_\parallel} Re\Big\{(\mu^2/8+i m)^2\Big[H_{-2}((\mu^2/8+i m)R_\parallel)-Y_{2}((\mu^2/8+i m)R_\parallel)\Big] \Big\},
    \end{split}
\end{equation}
where $\mathbf{R}_\parallel = \mathbf{b}_{\parallel 1} - \mathbf{b}_{\parallel 2}$ denotes the in-plane distance vector between the two sources, with $R_\parallel = |\mathbf{R}_\parallel|$ representing its magnitude. The subscript $V^{(1)}V^{(2)}$  refers to the interaction energy between the two topological sources in the planar gauge sector.

Given the complexity of the expression in (\ref{etoptop}), we divide our analysis into two distinct cases: the time-like case, where $\mathbf{V}^{(1)} = \mathbf{V}^{(2)} = 0$, and the space-like case, where $V^{(1)0} = V^{(2)0} = 0$.

\subsubsection{Time-like case}

For the time-like case, the interaction energy (\ref{etoptop}) becomes
\begin{equation}\label{eTTtemporal}
    \begin{split}
        &E_{V^{(1)}V^{(2)}}(m,\mu,R_\parallel)=\frac{V^{(1)0}V^{(2)0}}{4}Re\Big\{\frac{2}{R_\parallel}(\mu^2/8+i m)H_{-1}((\mu^2/8+i m)R_\parallel)\\
        &+(\mu^2/8+i m)^2\Big[Y_0((\mu^2/8+i m)R_\parallel)+H_{-2}((\mu^2/8+i m)R_\parallel)\Big] \Big\}.
    \end{split}
\end{equation}

We now present some remarks regarding the result above. In the temporal case, the topological source given by (\ref{sourceT}) is equivalent to a Dirac point, which can be interpreted as a transverse cross-section of a solenoid with zero radius but finite magnetic flux $\Phi$ \cite{EPJC802020,FABAAN2,DP3,LHCFABAFFEpjc}. This correspondence is established through the identification $V^{(i)0} = -\Phi^{(i)}$ (or more generally, $V^{(i)}_{\mu} = -\Phi^{(i)}\eta^{0}_{\parallel \mu}$), where $i=1,2$. Consequently, the interaction energy in (\ref{eTTtemporal}) can be understood as the interaction between two Dirac points, one located at $\mathbf{b}_{\parallel 1}$ with magnetic flux $\Phi^{(1)}$, and the other at $\mathbf{b}_{\parallel 2}$ with flux $\Phi^{(2)}$.

The imaginary argument in the Struve and Neumann functions in (\ref{eTTtemporal}) suppresses the oscillatory behavior, leading to an exponential-like decay, similar to the case of two stationary sources considered earlier. The interaction diverges when the topological sources (or, in this case, the Dirac points) overlap, but unlike the previous scenario, this divergence is not controlled by the coupling $\mu$ or the Chern-Simons mass $m$. In all cases, the interaction is repulsive when $V^{(1)0}$ and $V^{(2)0}$ have the same sign, and attractive when their signs differ.

In the decoupling limit ($\mu \to 0$), the interaction energy in Eq. (\ref{eTTtemporal}) arises purely from the Chern-Simons planar gauge field. Moreover, it vanishes in the limit $m \to 0$. In this regime, the quantities $m V^{(1)0}$ and $m V^{(2)0}$ can be interpreted as effective charges, and the time-like topological sources behave as two stationary point-like charges, as described in Eq. (\ref{classmaxchernint}). To make this correspondence explicit, setting $\mu = 0$ in Eq. (\ref{eTTtemporal}) yields
\begin{equation}
    \begin{split}
        E_{V^{(1)}V^{(2)}}(m,\mu=0,R_\parallel)&=\frac{V^{(1)0}V^{(2)0}m^2}{2\pi} K_{0}(m R_\parallel),\\
        &=\frac{\Big(m V^{(1)0}\Big)\Big(m V^{(2)0}\Big)}{2\pi}K_0(m R_\parallel),\\
        &=\frac{q_1 q_2}{2\pi}K_0(mR_\parallel).
    \end{split}
\end{equation}

However, due to the coupling with the photon field, governed by the parameter $\mu$, the interaction energy is not solely a consequence of Chern-Simons dynamics. In the massless limit ($m \to 0$), the interaction in Eq.~(\ref{eTTtemporal}) exhibits a strong dependence on $\mu$, and it vanishes both when the planar gauge field decouples from the photon field ($\mu = 0$) and in the strong coupling regime ($\mu \to \infty$), where the planar gauge field becomes non-propagating, as expected.

\begin{figure}[H]
    \centering
    \includegraphics[width=0.5\linewidth]{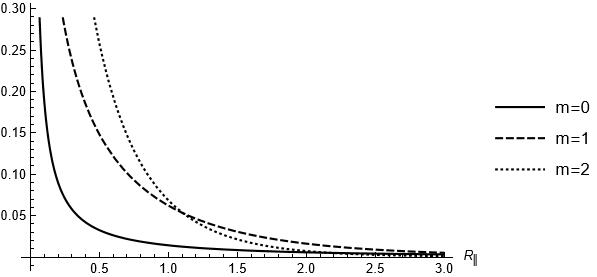}
    \caption{Plot of (\ref{eTTtemporal}), normalized by $V^{(1)0}V^{(2)0}$, as a function of $R_\parallel$ with fixed $\mu=1$ and: $m=0$ (solid line), $m=1$ (dashed line) and $m=2$ (dotted line).}
    \label{eTTtemporalmassas}
\end{figure}

\begin{figure}[H]
    \centering
    \includegraphics[width=0.5\linewidth]{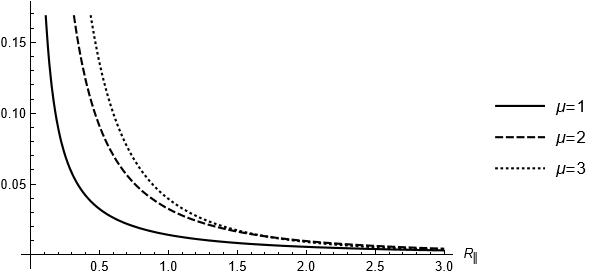}
    \caption{Plot of (\ref{eTTtemporal}), normalized by $V^{(1)0}V^{(2)0}$, as a function of $R_\parallel$ in the massless scenario $m=0$ with: $\mu=1$ (solid line), $\mu=2$ (dashed line) and $\mu=3$ (dotted line).}
    \label{eTTtemporalmassazero}
\end{figure}

To illustrate the discussion above, Figures (\ref{eTTtemporalmassas}) and (\ref{eTTtemporalmassazero}) show the interaction energy in Eq. (\ref{eTTtemporal}), normalized by $V^{(1)0}V^{(2)0}$, as a function of $R_\parallel$. In the first case, the coupling parameter $\mu$ is fixed, and different values of the mass $m$ are considered. In the second case, the massless scenario ($m = 0$) is analyzed for different values of the coupling parameter $\mu$.

\subsubsection{Space-like case}

In the spatial case, where $V^{(1)0}=V^{(2)0}=0$, the interaction energy given by (\ref{etoptop}) becomes
\begin{equation}\label{eTTespacial}
    \begin{split}
        &E_{V^{(1)}V^{(2)}}(m,\mu,R_\parallel)=\frac{\mathbf{V}^{(1)}.\mathbf{V}^{(2)}}{4R_\parallel}Re\Big\{(\mu^2/8+i m)\Big[2\Big(H_{-1}((\mu^2/8+i m)R_\parallel)\\
        &+2Y_{1}((\mu^2/8+i m)R_\parallel)\Big)-(\mu^2/8+i m)R_\parallel\Big(H_{-2}((\mu^2/8+i m)R_\parallel)\\
        &+Y_{0}((\mu^2/8+i m)R_\parallel)\Big) \Big]\Big\}+\Big(\mathbf{V}^{(1)}.\hat{\mathbf{R}}_\parallel\Big)\Big(\mathbf{V}^{(2)}.\hat{\mathbf{R}}_\parallel\Big) Re\Big\{(\mu^2/8+i m)^2\\
        &\times\Big[H_{-2}((\mu^2/8+i m)R_\parallel)-Y_{2}((\mu^2/8+i m)R_\parallel)\Big] \Big\},
    \end{split}
\end{equation}
where $\hat{\mathbf{R}}_\parallel = \mathbf{R}_\parallel / R_\parallel$ is a unit vector pointing from one topological source to the other. 

Let us now discuss the result above. Once again, due to the imaginary argument in the Struve and Neumann functions, the interaction in Eq. (\ref{eTTespacial}) diverges when the sources overlap and exhibits an exponential-like decay as $R_\parallel$ increases.

In the space-like case, the source in Eq.~(\ref{nontrivial}) corresponds to a planar electric dipole for the Chern--Simons gauge field \cite{EPJC802020}, with dipole moments given by $\mathbf{d}^{(i)} = \mathbf{V}^{(i)} \times \hat{n} = (V^{(i)2}, -V^{(i)1})$, (${\hat n}=(0,0,1)$ is the normal to the plate) located at positions $\mathbf{b}_{\parallel 1}$ and $\mathbf{b}_{\parallel 2}$, respectively. Accordingly, expression (\ref{eTTespacial}) represents the interaction energy between two electric dipoles in a $(2+1)$-dimensional theory with a massive gauge field, modified by its coupling to the photon field. Under the decoupling condition ($\mu = 0$), the interaction reproduces the result obtained for two electric dipoles in a $(2+1)$-dimensional massive gauge theory, namely,
\begin{equation}
    \begin{split}
        &E_{V^{(1)}V^{(2)}}(m,\mu=0,R_\parallel)=\frac{2m^2}{\pi}\Big(\mathbf{V}^{(1)}.\hat{\mathbf{R}}_\parallel\Big)\Big(\mathbf{V}^{(2)}.\hat{\mathbf{R}}_\parallel\Big) K_2(m R_\parallel)\\
        &-\frac{m\mathbf{V}^{(1)}.\mathbf{V}^{(2)}}{2\pi R_\parallel}\Big(m R_\parallel K_0(m R_\parallel)+4K_1(m R_\parallel)\Big),
    \end{split}
\end{equation}
\begin{equation}
    \begin{split}
        &E_{V^{(1)}V^{(2)}}(m=0,\mu=0,R_\parallel) = \frac{4 \Big(\mathbf{V}^{(1)}.\hat{\mathbf{R}}_\parallel\Big)\Big(\mathbf{V}^{(2)}.\hat{\mathbf{R}}_\parallel\Big) - 2 \mathbf{V}^{(1)}.\mathbf{V}^{(2)} }{\pi R^2}.
    \end{split}
\end{equation}

In order to gain a better understanding of the role played by the coupling parameter $\mu$ in the interaction between topological sources, let us consider the massless scenario, with the vectors $\mathbf{V}^{(1)}$ and $\mathbf{V}^{(2)}$ taken to be perpendicular to the separation vector $\mathbf{R}_\parallel$. In this setup, $\mathbf{V}^{(1)}$ and $\mathbf{V}^{(2)}$ are collinear, and the interaction is attractive or repulsive depending on whether their directions are opposite or the same, respectively. For this configuration, expression~(\ref{eTTespacial}) becomes
\begin{equation}\label{TTespecial}
    \begin{split}
        E_{V^{(1)}V^{(2)}}(m=0,\mu=0,R_\parallel)&=\frac{\mu^2\mathbf{V}^{(1)}.\mathbf{V}^{(2)}}{32R}\Big[4 Y_{1}(\mu^2 R/8)+2 H_{-1}(\mu^2 R/8)\\
        &-\frac{\mu^2R}{8}\Big(Y_0(\mu^2 R/8)+H_{-2}(\mu^2 R/8)\Big)\Big].
    \end{split}
\end{equation}

\begin{figure}[H]
    \centering
    \includegraphics[width=0.5\linewidth]{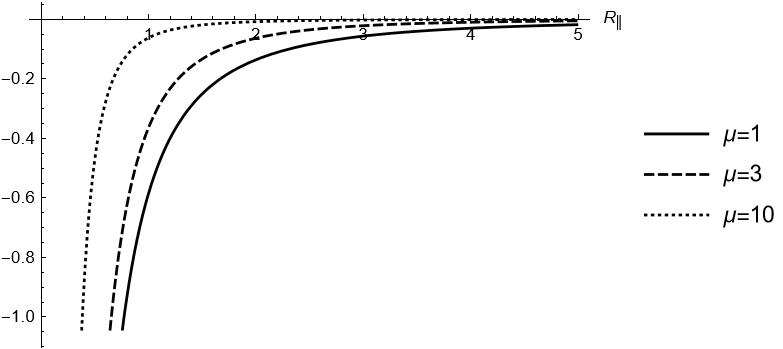}
    \caption{Plot of (\ref{TTespecial}), normalized by $\xi V^{(1)}V^{(2)}$, as a function of $R_\parallel$ with $\mu=1$ (solid line), $\mu=3$ (dashed line) and $\mu=10$ (dotted line).}
    \label{TTespecialPlot}
\end{figure}

The strength of the interaction in Eq. (\ref{TTespecial}) decreases as $\mu$ increases, exhibiting short-range behavior modulated by $\mu$, and vanishing entirely in the strong coupling limit. Moreover, in the configuration considered here, the interaction is attractive when $\mathbf{V}^{(1)}$ and $\mathbf{V}^{(2)}$ point in the same direction, and repulsive when they point in opposite directions. In Figure \ref{TTespecialPlot}, we plot Eq. (\ref{TTespecial}), normalized by $\xi V^{(1)} V^{(2)}$, as a function of $R_\parallel$ for several values of the coupling parameter.

The interaction energy in Eq. (\ref{eTTespacial}) also gives rise to a torque acting on the topological sources. To reveal this effect, let us consider the configuration where $\mathbf{V}^{(1)} = V^{(1)} \hat{x}$ and $\mathbf{V}^{(2)} = V^{(2)} \hat{y}$. We take source (1) to be fixed and calculate the torque on source (2). Let $\hat{\mathbf{R}}_\parallel = \cos(\theta) \hat{x} + \sin(\theta) \hat{y}$ be the unit vector pointing from source (1) to source (2). Substituting this setup into Eq. (\ref{eTTespacial}) and taking the negative derivative with respect to $\theta$, we obtain the torque $\boldsymbol{\mathcal{T}}_{V^{(1)} V^{(2)}}$ acting on source (2),
\begin{equation}
\label{torqueexp}
    \begin{split}
        &\mathbf{\mathcal{T}}_{V^{(1)}V^{(2)}}=-\frac{\partial}{\partial \theta}E_{V^{(1)}V^{(2)}}(m,\mu,R_\parallel)\;\hat{n},\\
        &=V^{(1)}V^{(2)}\Big(\sin{(\theta)}^2-\cos{(\theta)}^2\Big) Re\Big\{(\mu^2/8+i m)^2\\
        &\times\Big[H_{-2}((\mu^2/8+i m)R_\parallel)-Y_{2}((\mu^2/8+i m)R_\parallel)\Big]\Big\}\;\hat{n} ,
    \end{split}
\end{equation}
where $\hat n$ is the normal to the plate.

\begin{figure}[H]
    \centering
    \includegraphics[width=0.4\linewidth]{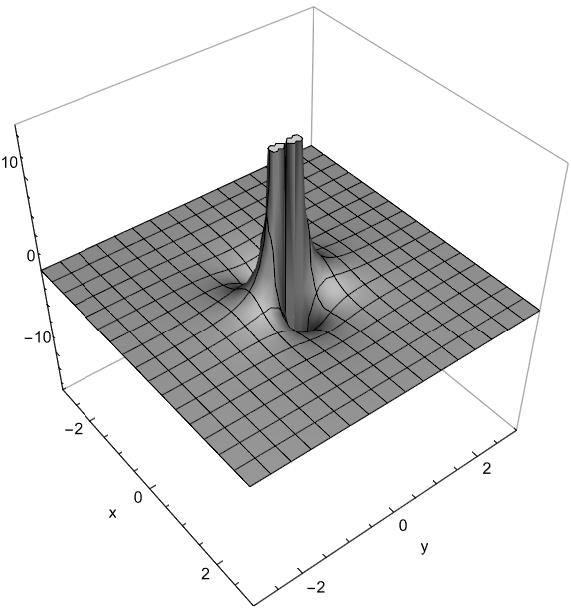}
    \caption{Plot of the component of (\ref{torqueexp}), normalized by $V^{(1)}V^{(2)}$, for $\mu=m=1$.}
    \label{torquetoptop}
\end{figure}

Figure \ref{torquetoptop} displays the torque component given in Eq. (\ref{torqueexp}), normalized by $V^{(1)} V^{(2)}$, as a function of the Cartesian coordinates $x$ and $y$. The origin of coordinates is set at the position $\mathbf{b}_{\parallel 1}$, and the plot corresponds to the case $\mu = m = 1$.

\section{Field Solutions for a Stationary Point-Like Charge: Highlighting Magnetoelectric Effects}
\label{V}

In this section, we analyze the field solutions for both the photon and the planar gauge fields produced by a stationary point-like electric charge.

We start by considering the classical solution for the matrix field in (\ref{campoecorrente})
\begin{eqnarray}
\mathbb{A}^{\mu}(x)=\int d^4y\;G^{\mu\nu}(x,y)\mathbb{J}_{\nu}(y)\ .
\end{eqnarray}
So, with the propagator (\ref{greencompleto}) at hand, we can compute the classical photon and planar gauge fields, which are given, respectively, by
\begin{equation}
\label{Avacuum}
    A_{\alpha}(x)=\sum_{i=1}^{2}\int d^4y\;G_{(1i) \alpha\nu}(x,y)\mathbb{J}_{(i)}^{\nu}(y),
\end{equation}
\begin{equation}
\label{ACALvacuum}
    \mathcal{A}_{\alpha}(x)=\sum_{i=1}^{2}\int d^4y\;G_{(2i) \alpha\nu}(x,y)\mathbb{J}_{(i)}^{\nu}(y).
\end{equation}

From equations (\ref{Avacuum}) and (\ref{ACALvacuum}), we observe that both $J$ and $\mathcal{J}$ act as sources for the two gauge fields simultaneously, due to the mixed components of the propagator.

Let us investigate the electric and magnetic fields in both domains generated by the source in (\ref{source1}). By substituting (\ref{source1}) and (\ref{greencompleto}) into equations (\ref{Avacuum}) and (\ref{ACALvacuum}), and then performing the integrations over $d\mathbf{y}$, $dy^0$, and $dp^0$ (in this order), we obtain the following integral expressions for the electric and magnetic fields associated with the photon and the planar gauge field, respectively
\begin{equation}
\label{photonelectric1}
    \begin{split}
        \mathbf{E}=\frac{Q\; \mathbf{x}}{4\pi |\mathbf{x}|^3}-\frac{\mu^2 Q}{32\pi}\int_0^\infty &dp_\parallel\;\frac{p_\parallel(p_\parallel+\mu^2/8)e^{-p_\parallel(R_\perp+|x^3-a|)}}{(p_\parallel+\mu^2/8)^2+m^2}\Bigg[J_{1}\big(p_\parallel|\mathbf{x}_\parallel-\mathbf{b}_\parallel|\big)\;\hat{x}_\parallel\\
        &+sign(x^3-a)J_{0}\big(p_\parallel|\mathbf{x}_\parallel-\mathbf{b}_\parallel|\big)\;\hat{n}\Bigg],
    \end{split}
\end{equation}
\begin{equation}\label{photonmag1}
    \begin{split}
        \mathbf{B}=-\frac{Q m \mu^2}{32\pi}\int_0^\infty &dp_\parallel\;\frac{p_\parallel e^{-p_\parallel(R_\perp+|x^3-a|)}}{(p_\parallel+\mu^2/8)^2+m^2}\Big[sign(x^3-a) \;J_1(p_\parallel|\mathbf{x}_\parallel-\mathbf{b}_\parallel|)\;\hat{x}_\parallel\\
        &+\Big(J_0(p_\parallel|\mathbf{x}_\parallel-\mathbf{b}_\parallel|)+\frac{J_1(p_\parallel|\mathbf{x}_\parallel-\mathbf{b}_\parallel|)}{p_\parallel x_\parallel}\Big)\hat{n}\Big],
    \end{split}
\end{equation}
\begin{equation}\label{planarelec1}
    \begin{split}
        \pmb{\mathcal{E}}=\frac{Q m \mu}{8\pi}\int_0^\infty dp_\parallel\;\frac{p_\parallel e^{-p_\parallel R_\perp}J_1(p_\parallel|\mathbf{x}_\parallel-\mathbf{b}_\parallel|)}{(p_\parallel+\mu^2/8)^2+m^2}\;\hat{x}_\parallel,
    \end{split}
\end{equation}
\begin{equation}\label{planarmag1}
    \begin{split}
        \mathcal{B}=-\frac{Q\mu}{8\pi |\mathbf{x}_\parallel-\mathbf{b}_\parallel|} \int_{0}^{\infty}&dp_\parallel\;\frac{(p_\parallel+\mu^2/8)e^{-p_\parallel R_\perp}}{(p_\parallel+\mu^2/8)^2+m^2}\Bigg[p_\parallel|\mathbf{x}_\parallel-\mathbf{b}_\parallel|J_{2}\big(p_\parallel|\mathbf{x}_\parallel-\mathbf{b}_\parallel|\big)\\
        &+J_{1}\big(p_\parallel|\mathbf{x}_\parallel-\mathbf{b}_\parallel|\big)\Bigg],
    \end{split}
\end{equation}
where $\hat{x}_\parallel=\mathbf{x}_\parallel/x_\parallel$, $R_\perp=|b^3-a|$ and $sign(x)$ is the sign function defined with $sign(x>0)=1$, $sign(x<0)=-1$ and $sign(x=0)=0$ .

Concerning the Maxwell sector, the electric field generated by the stationary point-like sources in (\ref{photonelectric1}) is given by the usual Coulomb field modified by a correction term that depends on the parameters of the planar gauge field, namely the Chern-Simons mass $m$ and the coupling parameter between the fields, $\mu$. This correction term involves a sign function that ensures this contribution points toward the plane. In the decoupling limit, the electric field reduces to the standard Coulomb field. Conversely, in the strong coupling regime, the resulting electric field mimics that of point-like sources in the presence of a perfectly conducting plate, thereby recovering the behavior predicted by the method of images.

The magnetic field in equation (\ref{photonmag1}) highlights the magnetoelectric property of the theory described by the Lagrangian in (\ref{lagrangeana2}). In this setup, a stationary point-like charge generates a magnetic field that strongly depends on the Chern-Simons mass $m$ of the planar gauge field and the interfield coupling parameter $\mu$. Notably, as already observed in the source-source interaction, the Chern-Simons mass appears not only in the denominator of the integrand but also in direct association with the electric charge density $Q$. This structure implies that the magnetoelectric effect vanishes when the Chern-Simons mass is zero. Furthermore, the magnetoelectric effect also disappears in both the decoupling and strong coupling limits. The parallel component of the magnetic field is proportional to the sign-function, which ensures a discontinuity across the plane.

Concerning the planar sector, the sources associated with the photon field also generate electric and magnetic fields, as shown in equation (\ref{ACALvacuum}). The electric field for the planar gauge field, generated by a stationary point-like charge, depends linearly on the Chern-Simons mass $m$ and the coupling parameter $\mu$, both appearing alongside the electric charge density $Q$. This planar electric field vanishes in the massless Chern-Simons limit, as well as in the decoupling and strong coupling regimes. This electric Chern-Simons field always points away from the projection of the source onto the plane.

The stationary point-like charge also generates a magnetic field in the planar sector, as shown in Eq. (\ref{planarmag1}), giving rise to a magnetoelectric effect within the plane. Unlike the Maxwell magnetic field, the Chern-Simons magnetic field does not vanish when the Chern-Simons mass is zero; here the mass acts as an inertia parameter that modulates the strength of the planar magnetic field. In the massless limit, only the coupling parameter $\mu$ appears, both in association with the electric charge and in the denominator of the integrand. As expected, the planar magnetic field vanishes in both the decoupling and strong coupling regimes.

\section{Conclusions}
\label{conclusoes}

In this work, we have proposed a field-theoretical description of a material layer with magnetoelectric properties within the framework of a planar gauge field. The dynamics of this field are governed by the Maxwell-Chern-Simons theory, where \( m \) denotes the Chern-Simons mass. It couples to the photon field through a Chern-Simons-type coupling. The interaction between the two gauge fields is mediated by a pseudoscalar parameter \( \mu \), which has dimensions of mass squared. We adopted the Feynman gauge for both fields.

The equations of motion have been derived for each field. In the photon sector, the presence of the planar gauge field induces a delta-function-type divergence, which can be interpreted in terms of polarization and magnetization defined along the material layer, both effectively described by the planar gauge field. However, instead of adopting this effective approach for the photon field, the full theory has been analyzed.

The propagator of the theory have been derived from a unified perspective by treating both gauge fields on equal footing, using a matrix formalism. This have involved rewriting the action in quadratic form and obtaining its inverse. The resulting matrix propagator does not correspond to a standard correction to the free gauge field; instead, it reflects modifications that require both fields to be considered simultaneously due to their mutual coupling. In the absence of this coupling, all interaction-induced corrections to the propagator vanish, and the propagator reduces to a diagonal form, describing the dynamics of two non-interacting gauge fields: the photon propagator and the planar Maxwell-Chern-Simons field, both in the Feynman gauge. Conversely, in the strong coupling limit, the off-diagonal elements of the propagator vanish. In this regime, the propagator in the electromagnetic sector reduces to the standard photon propagator in the Feynman gauge, modified by the presence of a perfectly conducting plate along the material layer. From the perspective of the planar gauge field, its dynamical terms reduce to a pure gauge contribution, with no physical observables arising from it. The coupling effectively modifies the mass of the planar gauge field, and in this limit, the field ceases to propagate.

Once the propagator has been derived, we have examined the interaction between a point-like charge in the photon sector and the planar gauge field. In the strong coupling regime, the magnetoelectric properties of the theory are suppressed, and the interaction reduces to the standard Coulomb force, independent of the Chern-Simons mass. Conversely, in the decoupling limit, the interaction vanishes entirely.

We have investigated the interaction between two static sources. As a first example, we have considered the interaction between two stationary point-like sources, both associated with the planar gauge field domain. In this case, the nature of the interaction follows the same principles as in electromagnetism: it is repulsive when the source strengths have the same sign and attractive when they have opposite signs. In the decoupling scenario (\( \mu = 0 \)), the interaction reduces to the standard result for stationary point-like sources in planar Maxwell-Chern-Simons electrodynamics. In the strong coupling regime, the planar gauge field ceases to propagate, and the interaction vanishes.

In the second example, we have examined the interaction energy between two stationary point-like sources from different domains: one associated with the photon sector and the other with the planar gauge field sector. In this context, the Chern-Simons mass plays the role of a coupling parameter for the interaction, even though it appears only in the dynamical equation for the planar gauge field and does not couple directly to the photon field. The interaction vanishes both in the massless limit (\( m = 0 \)) and in the strong mass limit (\( m \to \infty \)), as expected in the latter case.

The third configuration concerns the interaction between two topological sources, each defined by a constant three-dimensional pseudovector and coupled to the planar gauge field. These sources, which have recently gained attention in the literature, can emulate either a Dirac point, when the pseudovector has only time components, or an electric dipole, when it has only spatial components, depending on the specific configuration. The resulting interaction energy is anisotropic, which leads to a torque on the system. In the decoupling limit, where the sources effectively behave as Dirac points, the interaction reduces to that between two stationary point-like charges.

We have also examined the field solutions, both in the Maxwell sector and in the planar sector, induced by the presence of a stationary point-like electric charge. We have shown that both an electric and a magnetic field arise in the Maxwell sector, which highlights the magnetoelectric nature of the model. Similarly, both electric and magnetic fields are present in the planar sector. In the strong coupling limit ($\mu\to\infty$), the fields of the Chern-Simons sector vanish, as does the magnetic field of the Maxwell sector. In this regime, the electric field of the Maxwell sector coincides with that of a point charge near a perfectly conducting plate. When the Chern-Simons mass vanishes, there is no magnetic field in the Maxwell sector and no electric field in the planar sector. In the strong mass limit, all Chern-Simons fields vanish, along with the Maxwell magnetic field.

We leave as open questions the investigation of the model in other contexts, such as the propagation of electromagnetic waves in both sectors, the Casimir effect, and the computation of the effective action for the Maxwell field, among others. We hope that the discussion presented in this work will draw attention to two-field models as effective tools for describing material surfaces.

\textbf{\medskip{}}

\textbf{Acknowledgments} For financial support, H.L. Oliveira thanks to CNPq under the grant 152576/2024-0, F.A. Barone thanks to CNPq, under the grant 313426/2021 and J.P. Ferreira thanks to CNPq.

\end{document}